\documentclass[graphicx, 12pt]{article}
\usepackage{indentfirst}

\usepackage{graphicx}
\usepackage{caption}
\usepackage{subfigure}

\begin{document}

\small
\hoffset=-1truecm
\voffset=-2truecm
\title{\bf The time delay in strong gravitational lensing
with Gauss-Bonnet correction}
\author{Jingyun Man \hspace {1cm} Hongbo Cheng\footnote
{E-mail address: hbcheng@ecust.edu.cn}\\
Department of Physics, East China University of Science and
Technology,\\ Shanghai 200237, China\\
The Shanghai Key Laboratory of Astrophysics, Shanghai 200234,
China}

\date{}
\maketitle

\begin{abstract}
The time delay between two relativistic images in the strong
gravitational lensing governed by Gauss-Bonnet gravity is studied.
We derive and calculate the expression of time delay due to the
Gauss-Bonnet coupling. It is shown that the time delay for two
images with larger space each other is longer. We also find that
the ratio of Gauss-Bonnet coefficient and the mass of
gravitational source changes in the region like
$\frac{\alpha}{M}\in[0,2)$. The time delay is divergent with
$\frac{\alpha}{M}\longrightarrow 2$.
\end{abstract}

\vspace{8cm} \hspace{1cm} PACS number(s): 04.70.Bw, 98.62.Sb\\
Keywords: gravitational lensing, gravity, time delay

The gravitational lensing is due to the deflection of
electromagnetic radiation in a gravitational field [1-4]. The
relations between the deflection angle and the properties of
gravitational source are integral forms according to general
relativity and are certainly difficult to be investigated in
detail. In order to show how the gravitational source deviates the
path of light definitely, the integral expressions should be
discussed further. The expressions can be expanded in the limiting
cases such as weak field approximation and strong field limit.
Historically the gravitational lensing in the weak limit was used
to test the general relativity, but this kind of approach can not
describe the phenomena like the high bending and looping of the
electromagnetic rays. When the light goes very close to a heavy
compact body, its deflection angle will become larger and an
infinite series of images will generate. Only the gravitational
lensing in the strong limit can be used to explore these phenomena
that the light rays wind one or more times around the black hole
before reaching to the observer while exhibits the nature of the
massive source. In the past years more efforts have been
contributed to the strong gravitational lensing [5-8]. It should
be pointed out that a new technique by Bozza et.al. was utilized
to find the position of the relativistic images and their
magnification [9, 10]. Under the strong field limit the integral
expression for deflection angle is discussed around the radius of
photon sphere which leads the deflection angle to be infinitive.
We can list that the strong gravitational lensing was applied in a
Schwarzschild black hole [8, 11], gravitational source with naked
singularities [12], a Reissner-Nordstrom black hole [13], a GMGHS
charged black hole [14], a spining black hole [15, 16], a
braneworld black hole [17, 18], an Einstein-Born-Infeld black hole
[19], a black hole in Brans-Dicke theory [20], a black hole with
Barriola-Vilenkin monopole [21, 22], a deformed Horava-Lifshitz
black hole [23] and a black hole with Gauss-Bonnet correction
[24], etc..

As a kind of higher-dimensional gravity, the Einstein-Gauss-Bonnet
gravity is of considerable interest motivated by developments in
string theory. The theory is also a special case of Lovelock's
theory of gravitation. In this gravity, there is a dominating
quantum correction to classical general relativity and the new
term arises naturally in the low-energy limit of heterotic
superstring theory. The Gauss-Bonnet term appears as quadratic in
the curvature of the spacetime in the Lagrangian and certainly
regularizes the spacetime metric significantly. Up till now, both
qualitatively and quantitatively the Gauss-Bonnet coupling has not
been limited, so we can not settle for the more accurate
estimation on this correction. The Einstein-Gauss-Bonnet gravity
can be explored in different directions. Instead we are able to
describe the influence from Gauss-Bonnet term on the conclusions
in many kinds of important models [24-30].

It is necessary to research on the strong gravitational lensing in
the Schwarzschild black hole involving the Gauss-Bonnet
correction. In the gravitational lensing we should make
description of light's deviations like angular deflection and time
delay etc.. In a five-dimensional spacetime governed by
Gauss-Bonnet gravity, the deflection angle with logarithmic term,
corresponding parameters $\overline{a}$ and $\overline{b}$ and
some properties of relativistic images denoted as
$\theta_{\infty}$, $s$ and $r_{m}$ were derived and estimated in
the strong field limit in Ref. [24]. It is shown that the
Gauss-Bonnet term affects the parameters which could be detected
by astronomical instruments. It is also important to determine the
Gauss-Bonnet term's effect on time delay. The time delay is an
important window to explore the gravitational lensing system. In
the context of strong gravitational lensing, the multiple images
are formed and the light-travel-time along light paths
corresponding to different images is not the same. These time
delay are dimensional observables in gravitational lensing
measurements. Their measurements are also useful to determine the
nature of gravitational lensing system. To our knowledge, little
contribution is made to estimate the time delay for images in the
massive source with Gauss-Bonnet correction. In this paper we are
going to compute the analytical expressions for time delay between
any two images caused by the lens within the context of
Gauss-Bonnet gravity under strong field limit. This analytical
work will exhibit the significantly larger effect subject to both
the black hole and the Gauss-Bonnet coupling. At first we
introduce the spacetime metric dominated by Gauss-Bonnet term. We
derive the time delay between images caused by the
Gauss-Bonnet-corrected black hole in the case of strong field. We
calculate and plot the time delay associated with the Gauss-Bonnet
coupling. We summarize our results in the end.

The spherical metric describing the background of massive body
under the Gauss-Bonnet influence is given by [31],

\begin{equation}
ds^{2}=f(r)dt^{2}-\frac{dr^{2}}{f(r)}-r^{2}d\Omega_{3}^{2}
\end{equation}

\noindent with the help of action of Einstein-Gauss-Bonnet gravity
with five dimensions as follow,

\begin{equation}
I=\frac{1}{16\pi G_{5}}\int d^{5}x\sqrt{-g}[R+\frac{\alpha}{2}
(R^{2}-4R_{ab}R^{ab}+R_{abcd}R^{abcd})]
\end{equation}

\noindent where $R$, $R_{ab}$ and $R_{abcd}$ are Ricci scalar,
Ricci tensor and Riemann tensor respectively. $\alpha$ is
Gauss-Bonnet coefficient. $G_{5}$ is five-dimensional Newton's
constant. The component of metric (1) is,

\begin{equation}
f(r)=1+\frac{r^{2}}{2\alpha}(1-\sqrt{1+\frac{8\alpha M}{r^{4}}})
\end{equation}

\noindent Here we choose $G=c=1$. $M$ is subject to ADM mass. With
$f(r)=0$, the horizon radius of the black hole is,

\begin{equation}
r_{h}=\sqrt{2M-\alpha}
\end{equation}

In the five-dimensional spacetime, we set coordinates $x^{a}=(t,
r, \theta, \varphi, \psi)$. Here we let that both the observer and
the gravitational source lie in the equatorial plane with
condition $\theta=\frac{\pi}{2}$ for simplicity. According to Ref.
[32], the time difference between two photons travelling on
different trajectories is expressed as,

\begin{equation}
T_{1}-T_{2}=\widetilde{T}_{1}-\widetilde{T}_{2}+2\int_{r_{01}}^{r_{02}}
\frac{1}{f}dx
\end{equation}

\noindent where

\begin{equation}
\widetilde{T}(r_{0})=2\int_{r_{0}}^{\infty}[\frac{x\sqrt{f_{0}}}
{\sqrt{x^{2}f_{0}-x_{0}^{2}f}}-1]\frac{1}{f}dx
\end{equation}

\noindent We introduce the dimensionless variable
$x=\frac{r}{\sqrt{2M}}$ and $x_{0}=\frac{r_{0}}{\sqrt{2M}}$. Here
$r_{0}$ represents the minimum distance from the photon trajectory
to the gravitational source. The metric component
$f_{0}=f(r_{0})$. It should be pointed out that
$\widetilde{T}_{1}=\widetilde{T}(r_{01})$ and
$\widetilde{T}_{2}=\widetilde{T}(r_{02})$. Now $r_{01}$ and
$r_{02}$ are $r_{0}$ of two photons respectively. We denote the
time duration for the light ray to wind around the gravitational
source in the strong field limit,

\begin{equation}
\widetilde{T}(r_{0})=-\widetilde{A}\ln(\frac{u}{u_{m}}-1),
+\widetilde{B}
\end{equation}

\noindent where

\begin{equation}
\widetilde{A}=\sqrt{\frac{\sqrt{2}a}{(\sqrt{2}-\sqrt{2-a})(2-a)}},
\end{equation}

\begin{equation}
\widetilde{B}=\widetilde{A}\ln\frac{2a}{\sqrt{2-a}
(\sqrt{2}-\sqrt{2-a})}-\left(1.5649+0.8623a+0.2497a^{2}+O(a^{3})\right),
\end{equation}

\noindent and here $a=\frac{\alpha}{M}$, a dimensionless coupling.
Consider the formula between the impact parameter and the strong
coefficients of the deflection angle [9],

\begin{equation}
\frac{u}{u_{m}}-1=exp\left(\frac{\bar{B}-2n\pi\pm\gamma}{\bar{A}}\right),
\end{equation}

\noindent where (see [24])

\begin{equation}
\bar{A}=\frac{1}{\sqrt{2-a}},
\end{equation}

\noindent and

\begin{equation}
\bar{B}=-0.691-0.242a-0.104a^{2}+O(a^{3}).
\end{equation}

\noindent Here the impact parameter $u$ stands for the distance
from the lens to the null geodesic at the source position for
every each photon, and can be expressed at the closet approach as

\begin{equation}
u(x_{0})=\frac{x_{0}}{\sqrt{1+\frac{x_{0}^{2}}{a}\left(1-\sqrt{1+\frac{2a}{x_{0}^{4}}}\right)}}.
\end{equation}

\noindent The minimum impact parameter therefore become
$u_{m}=u(x_{m})$. We expand the $u$ to find the relation between
it and dimensionless variable $x$, then Eq. (10) can be rewritten
as

\begin{equation}
x_{0}=x_{m}+x_{m}\sqrt{\frac{2(\sqrt{2}-\sqrt{2-a})}{a\sqrt{2-1}}}exp\left(\frac{\bar{B}-2n\pi\pm\gamma}{\bar{A}}\right).
\end{equation}

Finally, according to Ea. (5), we obtain the time delay that the two images lie on the same side of the lens,

\begin{eqnarray}
\Delta T_{n,m}^{S}=2\pi(n-m)u_{m}
+\tilde{C}e^{\frac{\bar{B}}{2\bar{A}}}[\exp(-\frac{2m\pi\mp\gamma}{2\bar{A}})
-\exp(-\frac{2n\pi\mp\gamma}{2\bar{A}})],
\end{eqnarray}

\noindent and the time lag measuring images on opposite side of source

\begin{eqnarray}
\Delta T_{n,m}^{O}=2(\pi(n-m)u_{m}-\gamma)
+\tilde{C}e^{\frac{\bar{B}}{2\bar{A}}}[\exp(\frac{-2m\pi+\gamma}{2\bar{A}})
-\exp(\frac{-2n\pi-\gamma}{2\bar{A}})],
\end{eqnarray}

\noindent where

\begin{equation}
\tilde{C}=2^{\frac{7}{4}}\sqrt{\frac{a}{(2-a)(\sqrt{2}-\sqrt{2-a})}}.
\end{equation}

\noindent Here $\gamma$ is the angular separation between the
heavy compact body and the optical axis as seen from the lens. In
Eq. (15), the negative sign means that the two images are on the
same side of the lens and the positive one for the images standing
on the other side. The relation among the strong-deflection-angle
coefficient, the coefficient of time delay and the minimum impact
parameter,

\begin{equation}
u_{m}=\frac{\widetilde{A}}{\bar{A}},
\end{equation}

\noindent has already been used. More commonly, if the lens is
fortuitously aligned with the source galaxy, the massive objects
such as the huge dark matter concentrations in clusters of
galaxies or black holes can create the large bending angle and
multiple images. So, a extremely tiny angular separation is
reasonable like $\gamma\longrightarrow 0$, then $\Delta
T_{n,m}^{S}=\Delta T_{n,m}^{O}=\Delta T_{n,m}$. It is significant
that the time difference between two images depends on the
Gauss-Bonnet coupling, which can help us to detect how the
Gauss-Bonnet term corrects the general relativity.

As $a\longrightarrow 2$, the asymptotic behaviour of the time
delay is,

\begin{equation}
\lim_{a\longrightarrow 2}\Delta T_{n,m}=\infty,
\end{equation}

\noindent which corresponding to the divergent deflection angle in
Ref. [24]. We investigate the extreme value of time delay like
$\frac{d}{da}\Delta T_{n,m}|_{a=a_{0}}=0$, leading the equation
that the dimensionless variable $a_{0}$ obeys. Solving the
equation numerically, we list the estimations of $a_{0}$ in Table
1. Further we perform the burden calculation to find that,

\begin{equation}
\frac{d^{2}}{da^{2}}\Delta T_{n,m}|_{a=a_{0}}>0,
\end{equation}

\noindent which means that there exists a minimum value of time
delay as a function of $a=\frac{\alpha}{M}$ within the region
$a\in[0, 2)$.

\begin{table}[h!]
\begin{center}
\begin{tabular}{c||cccc|ccc|cc|c}\hline
 &\multicolumn{4}{c}{n-m=1}&\multicolumn{3}{c}{n-m=2}&\multicolumn{2}{c}{n-m=3}&\multicolumn{1}{c}{n-m=4}\\\hline
 $n,m$                                          & 2,1    & 3,2    & 4,3    & 5,4    & 3,1    & 4,2    & 5,3    & 4,1   & 5,2   & 5,1  \\ \hline
$a_{0}$                                         & 1.601  & 1.804  & 1.881  & 1.918  & 1.709  & 1.843  & 1.899  & 1.770 & 1.868 & 1.809 \\
min($\Delta T_{n,m}$)                           & 11.28  & 10.53  & 10.17  & 9.943  & 21.95  & 20.75  & 20.14  & 33.34 & 30.81 & 42.58 \\\hline
\end{tabular}
\caption{\label{tab:table1} The table shows the numerical
estimations for the minimum value of time delay between n-th and
m-th images and the correspongding $a_{0}$. }
\end{center}

\end{table}

 \begin{table}[h!]
 \begin{center}
 \begin{tabular}{c||cccccccccc}\hline

 $a$                                             & 0.01   & 0.1    & 1      & 1.6    & 1.7    & 1.8    & 1.9    & 1.99  & 1.999 & 1.9999 \\\hline
 $\frac{Second Term}{\Delta T_{2,1}}$ ($\%$)     & 0.23   & 0.26   & 1.10   & 5.22   & 7.60   & 11.8   & 20.7   & 45.0  & 54.3  & 57.2   \\
 $\frac{Second Term}{\Delta T_{3,1}}$ ($\%$)     & 0.11   & 0.13   & 0.57   & 3.03   & 4.62   & 7.72   & 15.2   & 41.5  & 53.1  & 56.9   \\\hline
 \end{tabular}
 \caption{\label{tab:table2} The percentage of the second term in time delay has been presented. }
 \end{center}

 \end{table}

In the strong field regime, a set of infinity relativistic images
will be produced due to different light paths. The n-order image
represents n laps a photon has circled. Now we plot the dependence
of time delay between the n-th and the m-th images on the
Gauss-Bonner coupling parameter $a$ in Figure. 1. The shapes of
all curves are similar, which implies that the Gauss-Bonnet
correction has general character reflected in the time lag between
two arbitrary images. From Figure. 1 and Table. 1, there exists a
minimum for each curve but the values of these minimums are
obviously distinguishable even for the time delay between images
with same difference of order, such as $min(\Delta
T_{2,1})>min(\Delta T_{3,2})$ for $n-m=1$. It is clear that the
more laps that two photons winds around the black hole differ, the
larger $\Delta T$ becomes, i.e., $\Delta T_{2,1}<\Delta T_{5,1}$.

When $a>a_{0}$, $\Delta T$ increases with dimensionless parameter
$a$ until a divergence occurs at $a=2$ due to the singular point
of space which leads a vanished event horizon or photon sphere. In
other words, as a function of Gauss-Bonnet coefficient, the time
lag between multiple images will be manifest if the Gauss-Bonnet
parameter is strong enough. It can be seen in Figure. 1 that once
$a$ exceeds the critical value $a_{0}$, the time delay grows
quickly, especially when $a\longrightarrow 2$. The increase of
function is offered from the second term of Eq. (15) or Eq. (16),
which is a subdominant term in the case of Schwarzschild black
hole [9]. And the percentage of the second term which is directly
related to the metric become more and more large (see Table. 2).
For instance, when $a=2-10^{-5}$ and $n=2$, $m=1$, the first term
in time delay contributes $8.90$ and the second term gives $12.36$
representing $58.2\%$ of the total time delay. Therefore, if the
action of Gauss-Bonnet do exist and its coefficient $\alpha$
approaches $2M$, we will receive a significant time delay by
observations of gravitational lensing and the effect from the
geometry which is dominated by $a$ will donate much enough to the
time lag to be distinguished upon the observational precision. We
can compare our results with the astrophysical measurement to
investigate the Gauss-Bonnet gravity.

In this paper we study the time delay between two relativistic
images in the strong gravitational lensing in the context of
Gauss-Bonnet gravity. We derive the analytical expression of time
delay between any two images to show the Gauss-Bonnet effect. The
time delay between the two images farther away from each other is
longer. When the Gauss-Bonnet coupling $\frac{\alpha}{M}$ exceeds
a critical value $a_{0}$, the time delay between two images will
fast increase and become the infinity at $a\longrightarrow 2$
which agrees to the divergent deflection angle. The appearance of
the Gauss-Bonnet coefficient enhances the metric related term in
time delay, which provides some new information about the geometry
due to the Gauss-Bonnet correction.

\vspace{1cm}
\noindent \textbf{Acknowledge}

This work is supported by NSFC No. 10875043 and is partly
supported by the Shanghai Research Foundation No. 07dz22020.

\newpage
\begin{figure}
\setlength{\belowcaptionskip}{10pt}\centering
\includegraphics[width=15cm]{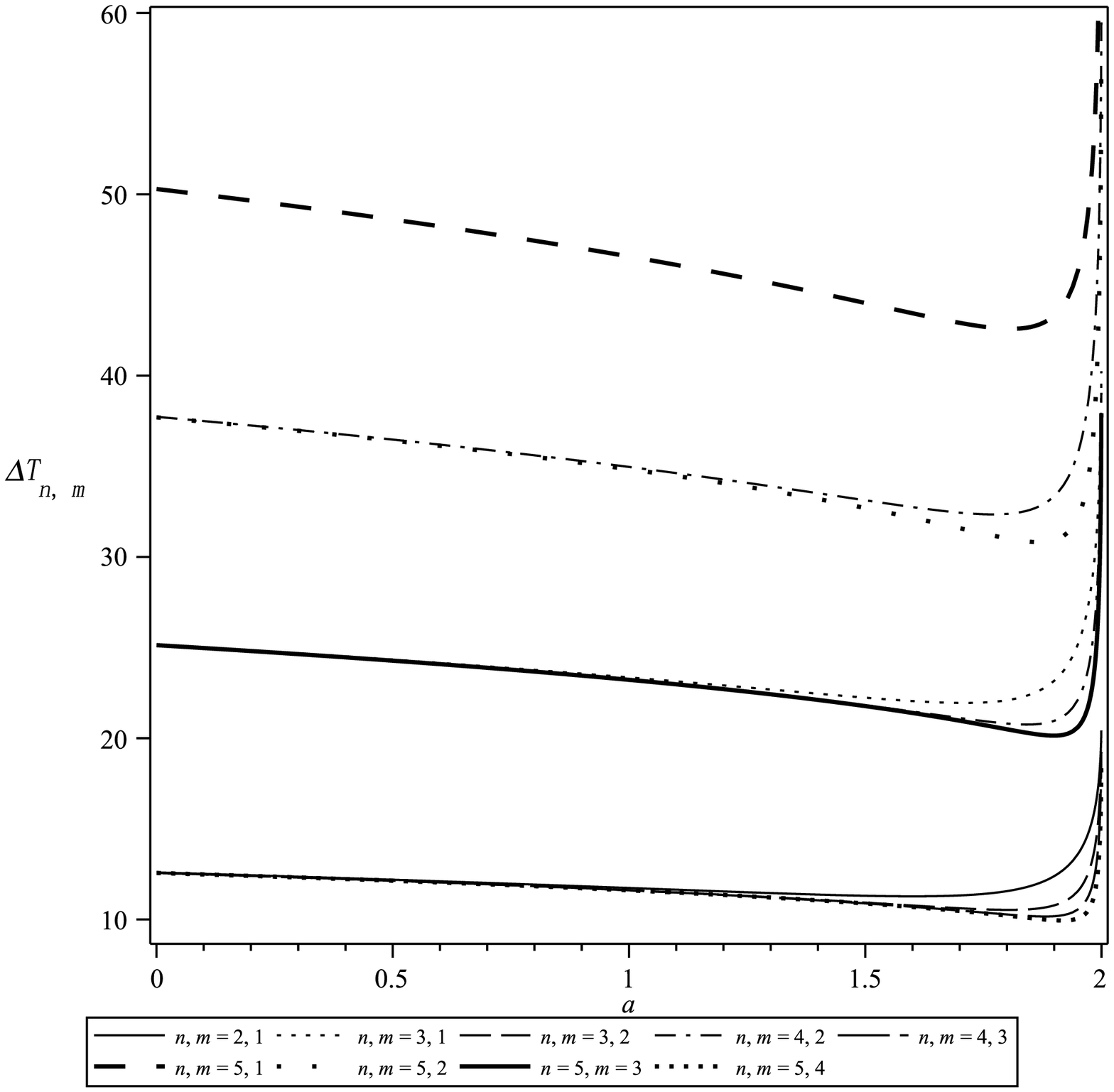}
\caption{The curves of time delay as a function of Gauss-Bonnet
coupling $a=\frac{\alpha}{M}$ limited by $a\in[0, 2)$.}
\end{figure}


\newpage
\begin{thebibliography}{99}
\bibitem{Schneider}P. Schneider, J. Ehlers, E. E. Falco,
Gravitational Lenses, Springer-Verlag, Berlin, 1992
\bibitem{Mollerach}S. Mollerach, E. Roulet, Gravitational Lensing
and Microlensing, World Scientific Publishing Co. Pte. Ltd. 2002
\bibitem{Blandford}R. D. Blandford, R. Narayan, Annu. Rev. Astron.
Astrophys. 30(1992)311
\bibitem{Refsdal}S. Refsdal, J. Surdej, Rep. Prog. Phys.
56(1994)117
\bibitem{Darwin}C. Darwin, Proc. Roy. Soc. Lond. 249(1959)180
\bibitem{Viergutz}S. U. Viergutz, Astron. Astrophys. 272(1993)355
\bibitem{Falcke}H. Falcke, F. Melia, E. Agol, Astrophys. J. Lett.
528(2000)L13
\bibitem{Virbhadra}K. S. Virbhadra, G. F. R. Ellis, Phys. Rev.
D62(2000)084003
\bibitem{Bozza}V. Bozza, S. Capozziello, G. Iovane, G. Scarpetta,
Gen. Rel. Grav. 33(2001)1535
\bibitem{Bozza}V. Bozza, Phys. Rev. D66(2002)103001
\bibitem{Frittelli}S. Frittelli, T. P. Kling, T. Newman, Phys.
Rev. D61(2000)064021
\bibitem{Virbhadra}K. S. Virbhadra, G. F. R. Ellis, Phys. Rev.
D65(2002)103004
\bibitem{Eiroa}E. F. Eiroa, G. E. Romero, D. F. Torres, Phys. Rev.
D66(2002)024010
\bibitem{Bhadra}A. Bhadra, Phys. Rev. D67(2003)103009
\bibitem{Bozza}V. Bozza, Phys. Rev. D67(2003)103006
\bibitem{Bozza}V. Bozza, F. Deluca, G. Scarpetta, M. Sereno, Phys.
Rev. D72(2005)083003
\bibitem{Whisker}R. Whisker, Phys. Rev. D71(2005)064004
\bibitem{Eiroa}E. F. Eiroa, Phys. Rev. D71(2005)083010
\bibitem{Eiroa}E. F. Eiroa, Phys. Rev. D73(2006)043002
\bibitem{Sarkar}K. Sarkar, A. Bhadra, Class. Quantum Grav.
23(2006)6101
\bibitem{Perlick}V. Perlick, Phys. Rev. D69(2004)064017
\bibitem{Cheng}H. Cheng, J. Man, Class. Quantum Grav.
28(2011)015001
\bibitem{Chen}S. Chen, J. Jing, Phys. Rev. D80(2009)024036
\bibitem{Sadeghi}J. Sadeghi, H. Vaez, JCAP1406(2014)028
\bibitem{Abdalla}E. Abdalla, R. A. Konoplya, Phys. Rev.
D72(2005)084006
\bibitem{Gleiser}R. J. Gleiser, G. Dotti, Phys. Rev.
D72(2005)124002
\bibitem{Sahabandu}C. Sahabandu, P. Suranyi, C. Vaz, L. C. R.
Wijewardhana, Phys. Rev. D73(2006)044009
\bibitem{Dominguez}A. E. Dominguez, E. Gallo, Phys. Rev.
D73(2006)064018
\bibitem{Maeda}H. Maeda, N. Dadhich, Phys. Rev. D74(2006)021501
\bibitem{Cheng}H. Cheng, Y. Liu, Chin. Phys. Lett. 25(2008)1160
\bibitem{Boulware}D. G. Boulware, S. Deser, Phys. Rev. Lett.
55(1985)2656
\bibitem{Weinberg}S. Weinberg, Gravitation and Cosmology:
Principles and Applications of the General Theory of Relativity,
Wiley, New York, 1972


\end{thebibliography}
\end{document}